\documentclass[a4paper,conference]{IEEEtran}
\newtheorem{definition}{Definition}
\newtheorem{remarks}{Remarks}
\IEEEoverridecommandlockouts
\usepackage{cite}
\usepackage{amsmath,amssymb,amsfonts}
\usepackage{algorithm2e}
\usepackage{graphicx}
\usepackage{textcomp}
\usepackage{xcolor}
\usepackage{amsmath}
\usepackage[skip=2pt,font=scriptsize]{caption}

\newtheorem{theorem}{Theorem}
\def\BibTeX{{\rm B\kern-.05em{\sc i\kern-.025em b}\kern-.08em
    T\kern-.1667em\lower.7ex\hbox{E}\kern-.125emX}}

\newcommand{\wblb}{\omega^b_{lb}}
\newcommand{\wbll}{\omega^l_{lb}}
\newcommand{\wbld}{\omega^d_{lb}}
\newcommand{\sdb}{\sigma_{db}}
\newcommand{\slb}{\sigma_{lb}}
\newcommand{\sld}{\sigma_{ld}}
\newcommand{\RR}{\mathbb{R}}
\newcommand{\RC}{\mathbb{R}^3}
\newcommand{\RM}{\mathbb{R}^{3 \times 3}}
\newcommand{\FF}{\mathcal{F}}

\begin{document}

\title{Linear Continuous Sliding Mode-based Attitude Controller with Modified Rodrigues Parameters Feedback\\
{\footnotesize}
\thanks{Final version. Published in 2020 International Conference on Radar, Antenna, Microwave, Electronics, and Telecommunications (ICRAMET)}
}

\author{\IEEEauthorblockN{Harry Septanto}
\IEEEauthorblockA{\textit{\textsuperscript{1}Satellite Technology Center}\\
\textit{National Institute of Aeronautics and Space (LAPAN)}\\
Bogor, Indonesia 
}
\IEEEauthorblockA{\textit{\textsuperscript{2}Ministry of Research and Technology/}\\
\textit{National Agency for Research and Innovation (BRIN)}\\
Jakarta, Indonesia\\
harry.septanto@lapan.go.id}
\and
\IEEEauthorblockN{Djoko Suprijanto}
\IEEEauthorblockA{\textit{Faculty of Mathematics and Natural Sciences} \\
\textit{Institute of Technology Bandung}\\
Bandung, Indonesia \\
djoko@math.itb.ac.id}

}

\maketitle

\begin{abstract}
This paper studies an attitude control system design based on modified Rodrigues parameters feedback. It employs a linear continuous sliding mode controller. The sliding mode controller is able to bring the existence of the sliding motion asymptotically. Besides, the attitude control system equilibrium point is proved to have an asymptotic stability guarantee through further analysis. This stability analysis is conducted since the sliding mode existence on the designed sliding surface does not imply the stability guarantee of the system’s equilibrium. This paper ends with some numerical examples that confirm the effectiveness of the designed attitude control system.
\end{abstract}


\section{Introduction}
Attitude control system design is a challenging problem. It contains nonlinear dynamics and kinematics models. Besides, the complexity increases since there is some kinematics \text{}representation with each own constraints. Research efforts have been conducted to design the attitude control
systems using various kinematics representations. For instance, Ozgoren \cite{Ozgoren2019} reported a comparison of the attitude control systems design based on Euler angles, quaternion, Euler angle-axis pairs, and orientation matrices (or rotation matrices \cite{Chaturvedi2011}). Another kinematics representation that has been also widely studied is modified Rodrigues parameters (MRP), e.g., \cite{Binette2014}, \cite{Doruk2009}, \cite{Tsiotras1998}, to name a few. MRP representation is not unique nor global \cite{Chaturvedi2011}, i.e., a physical orientation is not represented by a single-value in MRP as well as MRP represents not all physical orientations. Nevertheless, MRP can represent orientation between $-360^0$ and $360^0$ that cannot be conducted by Euler angles. Besides, its feedback control does not exhibit an unwinding phenomenon that may occur with quaternion representation \cite{Septanto2014}. Hence, this work is focused on the attitude control system design based on MRP kinematics representation.

Many approaches can be conducted in designing an attitude control system, .e.g., variable structure or sliding mode control approach. There are many types of sliding mode controller, e.g., discontinuous or switching \cite{Slotine1991}, terminal sliding mode \cite{Bhat1998}\cite{Venkataraman1993}, super-twisting \cite{Boiko2008}, saturation function-approximation \cite{Esfandiari1991}\cite{Brown2000}, and linear continuous-type of controller \cite{Zhou1992}\cite{Septanto2009} \cite{Santabudi2017}. The last type of sliding mode is less complex for implementation since it is a continuous system. Besides, it might chattering-free. The chattering occurs because switching of the control of the switching-type will excite the unmodeled dynamics \cite{Utkin2004}. Boiko and Fridman \cite{Boiko2005} showed the chattering of the so-called continuous sliding mode control systems. However, the analysis covered only the terminal sliding mode and super-twisting-type. This paper presents the resulting study of the sliding mode controller that is a linear continuous-type employed in the attitude control system using the MRP kinematics state feedback.

The paper is organized as follows. The next section describes the methodology of this research. Section III presents the main result that consist of a theorem, numerical simulations, and discussion. We end the paper by concluding remarks in Section IV.

\section{Methodology}

\subsection{Mathematical Preliminaries}

The time derivative of function $f$ is denoted by $\dot{f}$. A square matrix $G>0$ and $G<0$ mean a positive definite and a negative matrix, respectively. $I$ is the $3 \times 3$ identity matrix,  
$I = {\begin{bmatrix} 	1 & 0 & 0 \\	0 & 1 & 0 \\	0 & 0 & 1  \end{bmatrix}}$.

A vector of an angular velocity $\left( rad / s \right)$ are defined as follows: $\vec{\omega}_{lb} = {\wblb}^T \FF_b = {\omega^l_{lb}}^T \FF_l = {\omega^d_{lb}}^T \FF_d$, where ${\wblb, \wbll, \wbld \in\RC}$ and ${\FF_b, \FF_l, \FF_d \in \RM}$; $\wblb$ denotes the angular velocity of the satellite's body frame ($\FF_b$) with respect to the inertial reference frame ($\FF_l$) that is expressed in the body frame ($\FF_b$); $\omega^l_{lb}$ denotes the angular velocity of the satellite's body frame ($\FF_b$) with respect to the inertial reference frame ($\FF_l$) that is expressed in the inertial reference frame ($\FF_l$); and $\omega^d_{lb}$ denotes the angular velocity of the satellite's body frame ($\FF_b$) with respect to the inertial reference frame ($\FF_l$) that is expressed in the target or desired frame ($\FF_d$).
Note that the frame variables  $\FF_b, \FF_l$ and $\FF_d$ are  matrices whose rows consist of the vector basis of $\RC$.

\subsection{Rigid Body Dynamics and Kinematics}

The rigid body satellite dynamics with a $3 \times 3$ symmetric matrix of inertia calculated about its center of mass $J>0$ $\left( Kg\cdot m^2 \right)$, $J \in \RR^{3\times3}$, and the control torque expressed in the body frame $\tau$ $\left(N \cdot m \right)$, $\tau \in \RR^3$, is defined in \eqref{dynamics}. The attitude kinematics represented by modified Rodrigues parameters (MRP) is shown in \eqref{mrp1}-\eqref{mrp2} \cite{Tsiotras1998}, where \(\left[\sigma_{db}\right]^\times\) is a skew-symmetric matrix as defined in \eqref{skewsm}. Note that  $\sdb \in \RR^3$ denotes the attitude of the satellite's body frame ($\FF_b$) with respect to the desired frame ($\FF_d$),  where $\sigma_{db}=\slb -\sld$. It is also called the attitude error.

\begin{gather}  
J \dot{\omega}^b_{lb} = -\omega^b_{lb}\times J \omega^b_{lb} + \tau
\label{dynamics} 
\end{gather}

\begin{equation}  \label{mrp1} 
\dot{\sigma}_{db} = G\left( \sigma_{db} \right) \omega^b_{db}
\end{equation}

\begin{flushleft}
where
\end{flushleft}

\begin{equation} \label{mrp2} 
G\left(\sigma_{db}\right) = \frac{1}{2} \left(\frac{1-{\sigma_{db}}^T\sigma_{db}}{2}I-\left[\sigma_{db}\right]^\times + \sigma_{db} {\sigma_{db}}^T \right) 
\end{equation} 

\begin{flushleft}
and
\end{flushleft}

\begin{gather} 
\left[\sdb\right]^\times = 
{\begin{bmatrix} 
0 & -{\sdb}_3 & {\sdb}_2 \\
{\sdb}_3 & 0 & -{\sdb}_1 \\
-{\sdb}_2 & {\sdb}_1 & 0  \end{bmatrix}},
\sdb ={ \begin{bmatrix} 
{\sdb}_1\\
{\sdb}_2\\
{\sdb}_3
\end{bmatrix} } \label{skewsm} 
\end{gather}

\begin{definition} \label{def_SM}
Consider the rigid body dynamics \eqref{dynamics} and kinematics \eqref{mrp1}-\eqref{skewsm}. $\tau$ is an MRP-based feedback sliding mode controller if it is able to bring the system's states to reach the sliding surface asymptotically and to asymptotically stabilize the system's equilibrium point.
\end{definition}

\begin{remarks}
Some other authors stated that a finite-time stable of the sliding surface is required for a sliding mode existence, as stated in\cite{Slotine1991} and\cite{Bernuau2014}, to name a few. However, this work refers to that the sliding surface's asymptotically stable condition is necessary for a sliding
mode existence. This necessary condition may be found in \cite{RaymondA1988} and \cite{Zhou1992}, to name a few.
\end{remarks}

\subsection{Problem Statement}
Considering the rigid body dynamics \eqref{dynamics} and the MRP rotation kinematics \eqref{mrp1}-\eqref{skewsm}, design an attitude control system using MRP feedback and a linear continuous-type sliding mode controller for every constant desired attitude $\sld$ and zero desired angular velocity $\omega^b_{ld}$ from any initial angular velocity $\wblb\left(0\right)$ and attitude $\slb\left(0\right)$.

\section{Main Result}

The main result of this paper is presented by Theorem \ref{theorem1}. 

\begin{theorem} \label{theorem1}
Consider the rigid body dynamics \eqref{dynamics} and the MRP rotation kinematics \eqref{mrp1}-\eqref{mrp2} with the initial angular velocity $\omega^b_{lb}\left(t=0\right) \in \RR^3$, the initial attitude error $\sigma_{db}\left(t=0\right)\in \RR^3$, the constant desired attitude $\sld \in \RR^3$, and the zero desired angular velocity $\omega^b_{ld} = {\begin{bmatrix} 0 & 0 & 0 \end{bmatrix}}^T$.

Then  we have \(\tau=u_{eq} + u_{N}\) is a sliding mode controller with the sliding surface \eqref{ssurface}, where  $u_{eq}$  and $u_{N}$ are presented in equation \eqref{u_eq} and \eqref{u_N}, respectively, for certain $k_1, k_2 \in \mathbb{R}$, such that $k_1k_2>0$ and $L>0, L \in \mathbb{R}^{3 \times 3}$.

\begin{gather}  
S=
\Bigg\{ {\begin{bmatrix} \omega^b_{lb}\\ \sigma_{db} \end{bmatrix} }:
k_1 \omega^b_{lb} + k_2 \sigma_{db} = \xi, 
\xi = 
{\begin{bmatrix}
 0 & 0 & 0
\end{bmatrix}}^T \Bigg\} \label{ssurface}
\end{gather}

\begin{gather} 
u_{eq} =  \left(\omega^b_{lb} \times J \omega^b_{lb}\right) - \frac{k_2}{k_1}JG\left( \sigma_{db} \right)\omega^b_{lb}, \nonumber\\
\forall \begin{bmatrix}
{\omega^b_{lb}} \\ 
{\sigma_{db}} \end{bmatrix} \in S \label{u_eq}
\end{gather}

\begin{gather} 
u_{N} = -\frac{1}{k_1}J L \xi , \forall\xi \neq {\begin{bmatrix} 0 & 0 & 0\end{bmatrix}}^T \label{u_N}
\end{gather}
 
\end{theorem}

\begin{IEEEproof}
The sliding mode controller will be designed to follow the equivalent control method. First, we have to find the $u_{eq}$. At this step, we assume that the sliding mode is exist. Therefore, at the sliding surface, the system satisfies the condition \eqref{xi=0}

\begin{gather} 
\xi = k_1 \omega^b_{lb} + k_2 \sigma_{db} = {\begin{bmatrix} 0 & 0 & 0\end{bmatrix}}^T \nonumber\\
\Rightarrow \dot{\xi} = k_1 \dot{\omega}^b_{lb} + k_2 \dot{\sigma}_{db} = {\begin{bmatrix} 0 & 0 & 0\end{bmatrix}}^T \label{xi=0}
\end{gather}

Substituting $\dot{\omega}^b_{lb}$ and $\dot{\sigma}_{db}$ in \eqref{xi=0} by equation \eqref{dynamics} and \eqref{mrp1}, respectively, hence equation \eqref{ueq_deriv} is satisfied.

\begin{gather} 
{\begin{bmatrix} 0 & 0 & 0 \end{bmatrix}}^T =
k_1\Big(J^{-1}\left(-\omega^b_{lb} \times J \omega^b_{lb}\right) + J^{-1} \tau \Big) + \nonumber \\
k_2 G \left( \sigma_{db} \right)\omega^b_{db} \nonumber \\
\Leftrightarrow  J^{-1}\left(-\omega^b_{lb} \times J \omega^b_{lb}\right) + J^{-1} \tau = \nonumber \\
-\frac{k_2}{k_1} G \left( \sigma_{db} \right)\omega^b_{lb},\text{where } \omega^b_{ld} = {\begin{bmatrix} 0 & 0 & 0 \end{bmatrix}}^T \nonumber \\
\Leftrightarrow \tau = \left(\omega^b_{lb} \times J \omega^b_{lb}\right) - \frac{k_2}{k_1}J G \left( \sigma_{db} \right)\omega^b_{lb}, \nonumber\\
\forall k_1 \neq 0 \text{ and } k_2 \neq 0 \label{ueq_deriv}
\end{gather} 

At this point, we have the control torque that will only work when the states reach the sliding surface, $u_{eq}$, as shown in \eqref{ueq_formal}.

\begin{gather} 
 \tau = u_{eq} = \left(\omega^b_{lb} \times J \omega^b_{lb}\right) -  
 \frac{k_2}{k_1}J G \left( \sigma_{db} \right)\omega^b_{lb},\nonumber \\ \forall {\begin{bmatrix} \omega^b_{lb}\\ \sigma_{db} \end{bmatrix}} \in S \label{ueq_formal}
\end{gather}

Next, we have to determine the part of the control torque that works to ensure the existence of the sliding mode, i.e.,\(u_N\). This control torque is derived through the Lyapunov stability theory using Lyapunov function candidate, a positive definite function $V=\frac{1}{2}\xi^T\xi$. The complete derivation is shown in \eqref{sm_exist}.

\begin{gather}  
V=\frac{1}{2}\xi^T\xi>0, \forall \xi \neq {\begin{bmatrix} 0 & 0 & 0 \end{bmatrix}}^T\nonumber\\
\Rightarrow \dot{V}=\xi^T\dot{\xi}=\xi^T\Big(k_1\dot{\omega}^b_{lb}+ k_2\dot{\sigma}_{db}\Big)\label{sm_exist}
\end{gather}

Substituting  by $\dot{\omega}^b_{lb}$ in \eqref{dynamics} and $\dot{\sigma}_{db}$ in \eqref{mrp1}-\eqref{mrp2}, we obtain \eqref{sm_exist1}.

\begin{gather}
\dot{V}=\xi^T\Big(k_1 J^{-1}\left(-\omega^b_{lb} \times J \omega^b_{lb}\right)+k_1 J^{-1}\left(u_{eq}+u_N\right)+ \nonumber \\
 k_2 G\left(\sigma_{db}\right)\omega^b_{lb}\Big) \Leftrightarrow \dot{V}=k_1\xi^T J^{-1} u_N  \label{sm_exist1}
\end{gather}

If we have $u_N$ as shown in \eqref{uN_formal}, then \eqref{sm_exist2} is satisfied. Hence, $V$ is a Lyapunov function.

\begin{gather}
u_N = -\frac{1}{k_1} J L \xi, \forall L>0 \label{uN_formal}
\end{gather}

\begin{gather}
 \dot{V}=-\xi^T L \xi <0\label{sm_exist2}
\end{gather}

This fact implies that the sliding surface is asymptotically stable. In other words, it proves that the sliding mode exists. Nevertheless, it is not the end of the proof since we also want to make sure that the system's states will also reach the equilibrium point at $t \rightarrow \infty$.

Let $\bar{V} = \frac{1}{2}\xi^T\xi + 2\bar{k} \log_e\left(1+{\sigma_{db}}^T\sigma_{db}\right), \forall \bar{k}>0$ is a candidate Lyapunov function. Recall $G\left(\sigma_{db}\right)$ in \eqref{mrp2} and since ${\sigma_{db}}^T{\sigma_{db}}^T{\sigma_{db}} = {\sigma_{db}}^T{\sigma_{db}}{\sigma_{db}}^T$ and ${\sigma_{db}}^T{\left[\sigma_{db}\right]}^\times = {\begin{bmatrix} 0 & 0 & 0 \end{bmatrix}}^T$, hence we have an MRP property shown in \eqref{MRPproperty}. Noting this fact, we can have the time derivation of $\bar{V}$ that is shown in \eqref{origin_reach}.

\begin{gather}
{\sigma_{db}}^TG\left(\sigma_{db}\right) = {\sigma_{db}}^T\frac{1}{2} \Big(\frac{1-{\sigma_{db}}^T\sigma_{db}}{2}I-\left[\sigma_{db}\right]^\times + \nonumber\\
\sigma_{db} {\sigma_{db}}^T \Big) = \frac{1}{4}{\sigma_{db}}^T\left(1+\sigma_{db}^T\sigma_{db}\right)\label{MRPproperty}
\end{gather}

\begin{gather}
\dot{\bar{V}} = -\xi^T L \xi + \bar{k}{\omega^b_{lb}}^T \sigma_{db} \nonumber\\
 \Leftrightarrow \dot{\bar{V}}=-\bigl({k_1}^2 {\omega^b_{lb}}^T L \omega^b_{lb} + 2{k_1}{k_2}{\omega^b_{lb}}^T L \sigma_{db}+\nonumber\\
 {k_2}^2 {\sigma_{db}}^T L \sigma_{db}\bigr)+\bar{k}{\omega^b_{lb}}^T \sigma_{db}\label{origin_reach}
\end{gather}

Therefore, if $2 k_1 k_2 L = \bar{k} I$, then $\dot{\bar{V}}<0$. This fact are stated in \eqref{origin_reach1}. 

\begin{gather}
\dot{\bar{V}} = -{k_1}^2 {\omega^b_{lb}}^T L \omega^b_{lb} - {k_2}^2 {\sigma_{db}}^T L \sigma_{db}<0, \nonumber\\
\text{ if } 2 k_1 k_2 L = \bar{k} I>0 \label{origin_reach1}
\end{gather}

Since, ${\bar{V}>0}$ and ${\dot{\bar{V}}<0}$ for any non-zero value of the state $\left({\omega^b_{lb}}, {\sigma_{db}} \right)$, hence the equilibrium point, $\left({\omega^b_{lb}}, {\sigma_{db}} \right)
 = 
{\begin{bmatrix} 0 & 0 & 0 & 0 & 0 & 0 \end{bmatrix}}^T$, is asymptotically stable for any initial angular velocity $\omega^b_{lb}\left(t=0\right)$, the initial attitude error $\sigma_{db}\left(t=0\right)$, and the desired angular velocity $\omega^b_{ld} = 
{\begin{bmatrix}
0 & 0 & 0                                                                                                                                                                                                                                                              \end{bmatrix}}^T$.

In addition, since $\bar{k}>0$ and $L>0$, it implies more strict condition regarding $k_1$ and $k_2$, i.e.,  $k_1k_2>0$. 
Note that since $\bar{k}$ is any positive value, hence $L$ can be any positive definite matrix as well as $k_1$ and $k_2$ can be any non-zero scalar such that $k_1 k_2>0$. It completes the proof.
\label{proof}
\end{IEEEproof}

\begin{remarks}
	The designed sliding mode controller $\tau$ is a continuous but not linear feedback. "Linear continuous" term used to name the type of the sliding mode is based on the control structure of the $u_N$.
\end{remarks}

\begin{figure}
	\centerline{\includegraphics[scale=0.44]{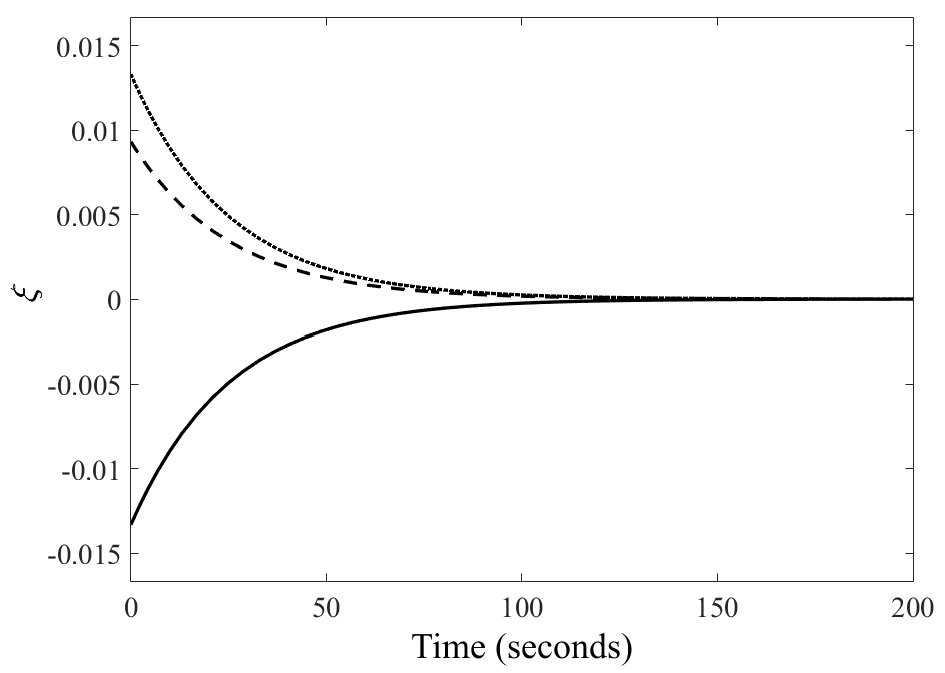}}
	\captionsetup{format=hang}
	\caption{Sliding motion; $\xi = \begin{bmatrix} \xi1 & \xi2 & \xi3 \end{bmatrix}^T; \xi1- ; \xi2--; \xi3\cdots $.}
	\label{fig0}
\end{figure}

\begin{figure}
	\centerline{\includegraphics[scale=0.44]{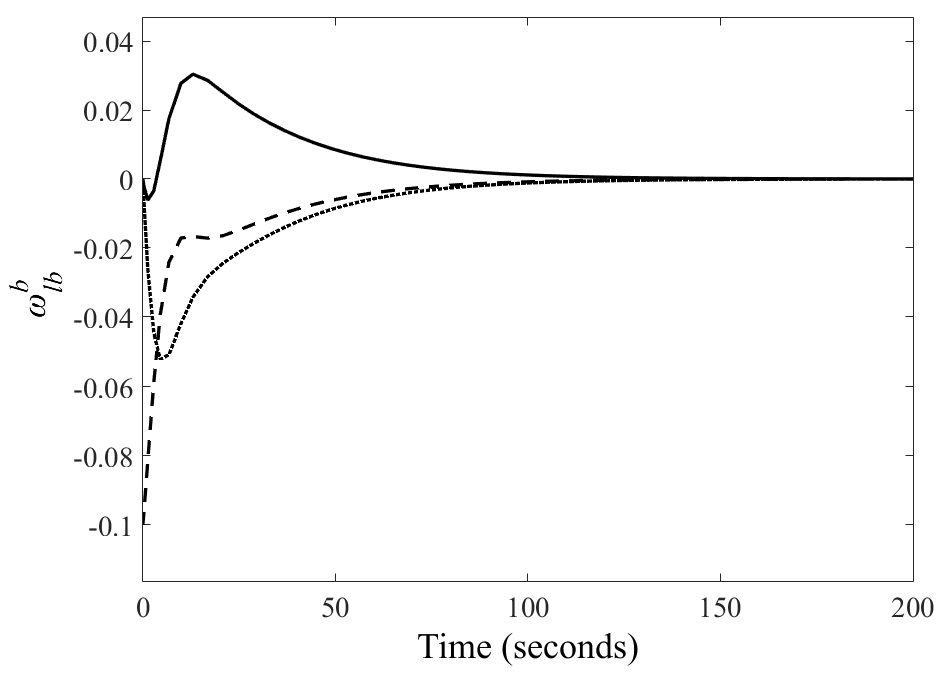}}
	\captionsetup{format=hang}
	\caption{Angular velocity of the satellite's body frame with respect to the inertial reference frame expressed in the satellite body frame; $\omega^b_{lb} =\begin{bmatrix} \omega^b_{lb1} & \omega^b_{lb2} & \omega^b_{lb3}\end{bmatrix}^T; \omega^b_{lb1}- ; \omega^b_{lb2}--; \omega^b_{lb3}\cdots$.}
	\label{fig1}
\end{figure}

\begin{figure}
	\centerline{\includegraphics[scale=0.44]{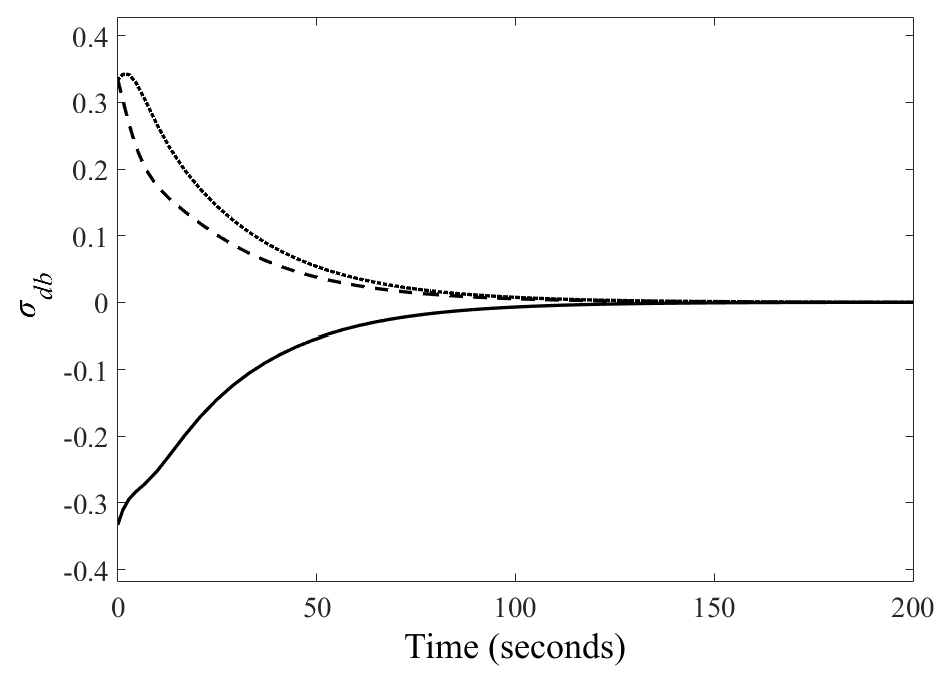}}
	\captionsetup{format=hang}
	\caption{Attitude of the satellite's body frame with respect to the desired frame (attitude error); $\sigma_{db} =\begin{bmatrix} \sigma_{db1} & \sigma_{db2} & \sigma_{db3} \end{bmatrix}^T; \sigma_{db1}- ; \sigma_{db2}{--}; \sigma_{db3}\cdots$.}
	\label{fig2}
\end{figure}

\begin{figure}
	\centerline{\includegraphics[scale=0.44]{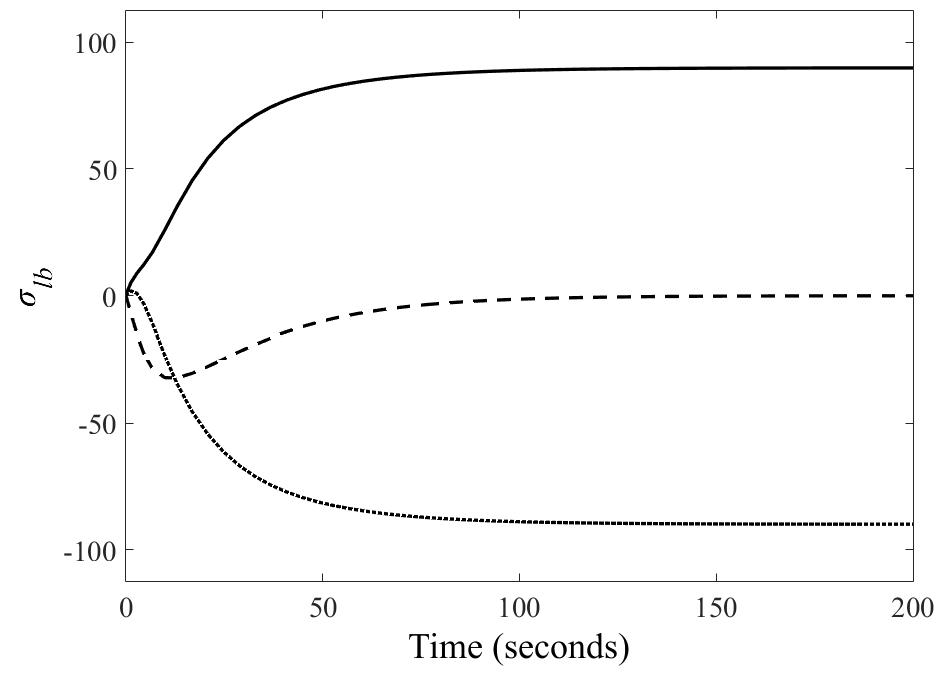}}
	\captionsetup{format=hang}
	\caption{Attitude of the satellite's body frame with respect to the inertial reference frame; $\sigma_{lb} =\begin{bmatrix} \sigma_{lb1} & \sigma_{lb2} & \sigma_{lb3} \end{bmatrix}^T; \sigma_{lb1}- ; \sigma_{lb2}--; \sigma_{lb3}\cdots$.}
	\label{fig3}
\end{figure}

\begin{figure}
	\centerline{\includegraphics[scale=0.44]{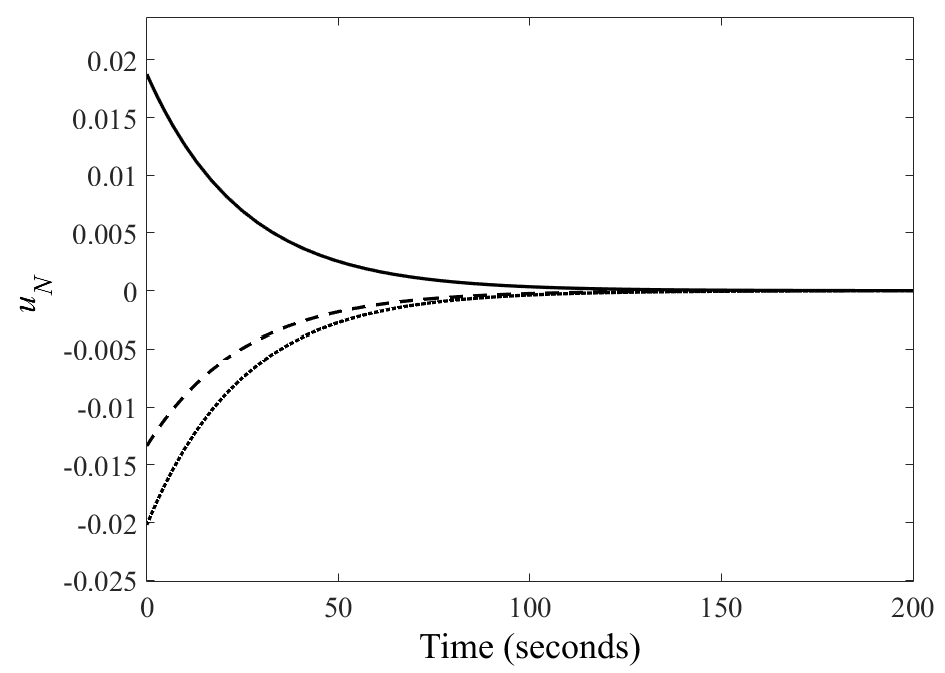}}
	\captionsetup{format=hang}
	\caption{Control signal $u_N$; ${u_N =\begin{bmatrix} u_{N1} & u_{N2} & u_{N3} \end{bmatrix}^T;}$ $u_{N1}- ;$ $u_{N2}{--};$ $u_{N3}\cdots$.}
	\label{fig4}
\end{figure}

\begin{figure}
	\centerline{\includegraphics[scale=0.44]{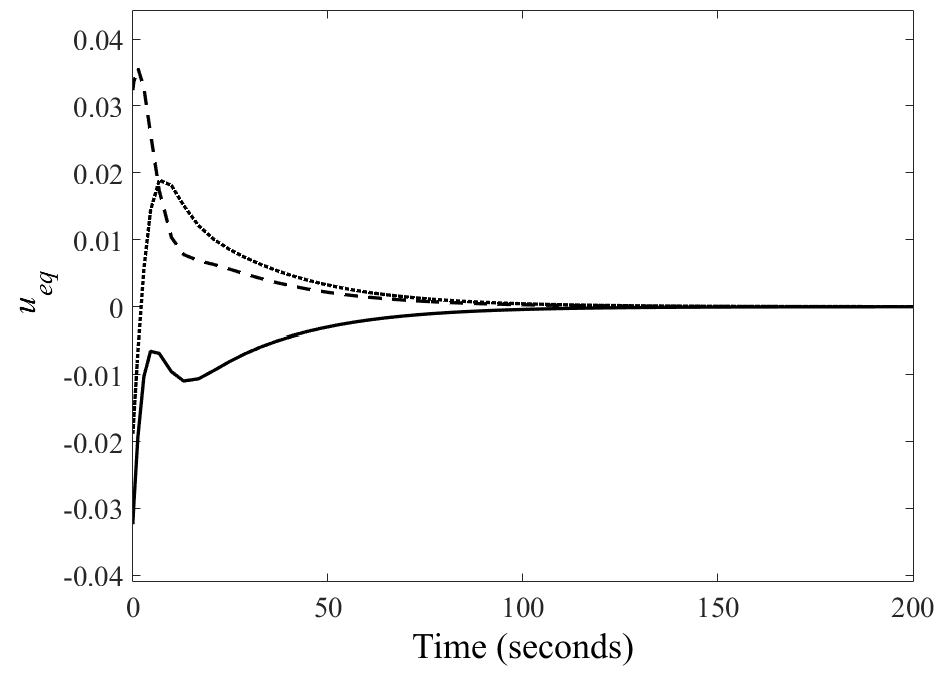}}
	\captionsetup{format=hang}
	\caption{Control signal $u_{eq}$; ${u_{eq} =\begin{bmatrix} u_{eq1} & u_{eq2} & u_{eq3} \end{bmatrix}^T;}$ $u_{eq1}- ;$ $u_{eq2}{--};$ $u_{eq3}\cdots$. }
	\label{fig5}
\end{figure}

Some numerical examples are presented to figure out how effective the designed sliding mode-based attitude control system is. The simulations are conducted using a variable-step solver. The rigid body satellite's moment of inertia, the controller parameters, and the initial conditions are presented in \eqref{simul1}, \eqref{simul2}, and \eqref{simul3}, respectively. The value of the moment of inertia $J$ is adopted from \cite{Septanto2017}, while the controller parameters are arbitrary chosen. The desired attitude $\sld$ value is taken from the simulation setting in \cite{Doruk2009}.

\begin{gather}
 J = {\begin{bmatrix} 
 1.49 & 0.054 & 0.0442 \\ 
 0.054 & 1.51 & 0\\ 
 0.0442 & 0 & 1.56
 \end{bmatrix}}
 \label{simul1}
\end{gather}

\begin{gather}
 k_1 = k_2 = 0.04, L = 0.04I
 \label{simul2}
\end{gather}

\begin{gather}
 \wblb\left(0\right) = {\begin{bmatrix} 0 & -0.1 & 0 \end{bmatrix}}^T,\nonumber\\
 \slb\left(0\right) = {\begin{bmatrix} 0 & 0 & 0 \end{bmatrix}}^T,\nonumber\\
 \sld = {\begin{bmatrix} 0.3333 & -0.3333 & -0.3333 \end{bmatrix}}^T
  \label{simul3}
\end{gather}

Fig.~\ref{fig0}-Fig.~\ref{fig5} show some dynamics characteristics relating to the simulation setting. Fig.~\ref{fig0} verifies that the sliding motion exists. Meanwhile, Fig.~\ref{fig4} and Fig.~\ref{fig5} confirm that the control signal $u_N$ and $u_{eq}$, respectively, are also able to bring the trajectories converge to the equilibrium (Fig.~\ref{fig1} and Fig.~\ref{fig2}), i.e., reach the desired states (Fig.~\ref{fig3}).

\section{Concluding Remarks}
The designed attitude control system using a linear continuous-type sliding mode controller with the attitude state feedback in modified Rodrigues parameters (MRP) representation has been presented. It guarantees the sliding mode to exist asymptotically. Furthermore, the equilibrium point of the control system has an asymptotic stability guarantee. This stability analysis is conducted since the sliding mode existence on the designed sliding surface does not imply the stability guarantee of the system’s equilibrium. Numerical examples verify that the sliding motion exists and the trajectories converge to the desired states. 

Additional future work would concentrate on investigating the robustness properties of this control system. Besides, the attitude controller's stability analysis in the discrete-time domain for its digital implementation would also be the future works.

\section*{Acknowledgments}

This research is supported by the Ministry of Research and Technology/ National Agency for Research and Innovation (BRIN), Jakarta, Republic of Indonesia. The authors also acknowledge the Satellite Technology Center, National Institute of Aeronautics and Space (LAPAN) for providing the research facilities. HS is the main contributor of this paper with detail contributions as follows: HS – idea, proof derivation, simulation, discussion, paper preparation; DS – discussion, reviewing, editing. 

\bibliographystyle{IEEEtran}
\bibliography{MRP}
\end{document}